%% file: main.tex
\documentclass[conference]{IEEEtran}

\usepackage{enumitem}
\usepackage{booktabs}
\usepackage{multirow}
\usepackage[skip=1pt]{caption}
\usepackage{seqsplit}
\usepackage{pgfplots}
\usepackage{float}
\usepackage{subcaption}
\usepackage[table]{xcolor}
\usepackage{tabularx}
\usepackage{pifont}
\usepackage{comment}
\usepackage{cite}
\pgfplotsset{compat=1.17}
\usepackage[hidelinks]{hyperref}

\newcommand{\tool}{\textsc{BlindSpot}}

\begin{document}

\title{\tool{}: Enabling Bystander-Controlled Privacy Signaling for Camera-Enabled Devices}

\author{
    \IEEEauthorblockN{Jad Al Aaraj}
    \IEEEauthorblockA{University of California, Irvine\\
    Irvine, CA, USA\\
    Email: jalaaraj@uci.edu}
    \and
    \IEEEauthorblockN{Athina Markopoulou}
    \IEEEauthorblockA{University of California, Irvine\\
    Irvine, CA, USA\\
    Email: athina@uci.edu}
}

\maketitle

\input{sections/abstract}

\begin{IEEEkeywords}
Privacy; Bystanders; Camera-Enabled Mobile Devices; Localization; Real-time; Face Detection; Gesture Recognition; Visual Light Communication (VLC); Ultra-Wideband (UWB).
\end{IEEEkeywords}

\input{sections/introduction}
\input{sections/motivation}
\input{sections/related_work}
\input{sections/sys_design}
\input{sections/evaluation}
\input{sections/discussion}
\input{sections/conclusion}
\input{sections/acks}

\bibliographystyle{IEEEtran}
\bibliography{main}

\input{sections/appendix}

\end{document}

%% file: sections/abstract.tex
\begin{abstract}
Camera-equipped mobile devices, such as phones, smart glasses, and AR headsets, pose a privacy challenge for bystanders, who currently lack effective real-time mechanisms to control the capture of their picture, video, including their face. 
We present \tool{}, an on-device system that enables bystanders to manage their own privacy by signaling their privacy preferences in real-time without previously sharing any sensitive information.
Our main contribution is the design and comparative evaluation of three distinct signaling modalities: a hand gesture mechanism, a significantly improved visible light communication (VLC) protocol, and a novel ultra-wideband (UWB) communication protocol.
For all these modalities, we also design a validation mechanism that uses geometric consistency checks to verify the origin of a signal relative to the sending bystander, and defend against impersonation attacks.
We implement the complete system (\tool{}) on a commodity smartphone and conduct a comprehensive evaluation of each modality's accuracy and latency across various distances, lighting conditions, and user movements. Our results demonstrate the feasibility  of these novel bystander signaling techniques 
and  their trade-offs  in terms of system performance and convenience.
\end{abstract}

%% file: sections/introduction.tex
\section{Introduction}
\label{sec:intro}

The proliferation of camera-enabled mobile and wearable devices has enabled the constant recording of our surroundings. This trend, accelerated by the rise of artificial intelligence (AI) applications on everything from smart glasses to everyday smartphones, violates the privacy of bystanders, {\em i.e.,} individuals who are present in the camera's field of view but have not consented to being recorded~\cite{corbett2023securing}. Bystanders currently lack an explicit and reliable mechanism to manage how their image or video is captured and used, in real-time, hindering the social acceptance of these powerful camera-enabled technologies~\cite{denning-2014, hoyle-2014}. In this paper, we focus on the privacy of bystanders' face, a sensitive biometric.

Designing a system that grants privacy control to individual bystanders presents several fundamental challenges. First, there is an {\em agency gap}: most state-of-the-art privacy controls are device-centric, requiring the owner of the camera-enabled device to mediate privacy on behalf of the bystander
\footnote{For example,  
a user's attention is used as a proxy for bystander consent in \cite{corbett2023bystandar}, which is not correct in public spaces. Consider a user asking a stranger for directions; the user's gaze would be misinterpreted as a sign of a trusted interaction, causing the system to leave the stranger's face unprotected without their consent. More critically, this model provides no direct channel for a bystander to signal their preference, leaving them powerless. 
 }.
In this work, we give agency to the bystanders, by providing them with an explicit mechanism  to signal and assert their own privacy choices.
Second, to the best of our knowledge, there is no current solution that provides {\em real-time} bystander-controlled protection.
 Systems that offload processing ({\em e.g.,} face blurring) to the cloud can introduce unacceptable latency and apply only after the video is shared with a remote server~\cite{aditya2016ipic}. 
 Third, some solutions often trade one threat model for another. For example, they require bystanders to register and upload their faceprints to a trusted database in-advance, so they can be identified and have their face blurred in real-time ~\cite{shu2018cardea}.
A practical system must be on-device, real-time, and biometric-free; navigating the tight computational and power constraints of mobile hardware; and addressing usability and cost.

To meet these requirements, we design {\bf \tool{}}, a novel system architecture that closes the agency gap and
enables bystanders to manage their privacy, by explicitly signaling their preferences to the user's device. The device then detects and tracks the bystander's face in the video, and blurs/unblurs it according to those preferences.
 
In order to enable signaling privacy preferences from the bystander to the camera-enabled device, we propose, implement and evaluate three possible options:
\begin{itemize}[nosep, leftmargin=*]
    \item \textbf{Modality 1: Hand gestures} for intuitive, hardware-free signaling. A hand gesture is device-free but can be ambiguous and prone to misinterpretation. We design and implement a velocity-based gesture recognizer that is robust to variable frame rates.
    
    \item \textbf{Modality 2: Visual Light Communication (VLC)} for signaling via a simple, low-cost LED, available as standalone pin (no phone required), or on wearables, such as smart glasses. The bystander, carrying the LED, can send an explicit signal to the camera-enabled device through a dedicated beacon.
    Our design utilizes a GPU-accelerated peak detection filter and a Viterbi-based decoding algorithm to achieve robust decoding at range and under motion.
    
    \item \textbf{Modality 3: Ultra-Wideband (UWB) tag}: a radio technology designed for
    precise indoor localization and now integrated into modern smartphones ({\em e.g.,} iPhone). This RF-based approach is robust to visual interference but can introduce significant power overhead. We design a hybrid UWB and Bluetooth (BLE) protocol that minimizes power consumption by avoiding continuous UWB ranging.
\end{itemize}

 \tool{}'s current design uses one signaling modality at a time.
We argue that no single approach can adequately solve the bystander privacy problem in all settings, due to their different characteristics, summarized in Table~\ref{tab:tradeoffs}.
Furthermore, we consider a bystander impersonation attack, where an adversary attempts to signal on a victim's behalf, and we propose a defense that validates the proximity of the signal's origin to  the bystander's face.

We implemented and evaluated \tool{} on a Google Pixel 8 Pro. Our evaluation demonstrates the effectiveness and robustness of each modality in real-world scenarios. We quantify the distinct trade-offs each presents: the gesture modality is device-free and offers low latency ($\sim$160\,ms) but is limited to a 3m range and incurs high CPU overhead. In contrast, the UWB modality is robust to all lighting conditions and scales to multiple users with near-constant latency ($\sim$2000\,ms) and minimal CPU cost, at the expense of requiring dedicated hardware. The VLC modality offers a balance between the two. A preliminary user study confirms the value of our multi-modal design and reveals a strong user preference for the discretion afforded by the UWB approach.

 In summary,  our novel contributions include both the overall \tool{}~ system; and the design, implementation, and evaluation of three novel signaling methods, built on top of each of the three modalities, specifically for bystander privacy.  To the best of our knowledge, modalities 1 and 3 have not been implemented before for bystander privacy; modality 2 improves upon our own prior work named BystandARIA~\cite{alaaraj2025bystandaria}.
 \tool{} is the first to achieve bystander-driven signaling of privacy preferences, in real-time, and without reliance on the cloud for matching faceprints. Furthermore, its modular design makes it an extensible framework for bystander-controlled privacy on commodity mobile devices.

The remainder of the paper is organized as follows. Section~\ref{sec:motivation} defines the problem setting and terminology. Section~\ref{sec:related} reviews related work. Section~\ref{sec:design} details the system architecture and the design of each modality. Section~\ref{sec:evaluation} presents our experimental evaluation. Section~\ref{sec:discussion} discusses the findings, recommendations, future directions and ethical considerations. Section~\ref{sec:conclusion} concludes the paper.

\begin{table}[t]
\centering
\caption{Characteristics of Signaling Modalities.}
\label{tab:tradeoffs}
\resizebox{\columnwidth}{!}{%
\begin{tabular}{@{}lcccc@{}}
\toprule
\multicolumn{2}{c}{\textbf{Property}} & \textbf{Gesture} & \textbf{VLC (LED)} & \textbf{UWB} \\ \midrule
\multirow{4}{*}{\rotatebox[origin=c]{90}{\parbox{1.5cm}{\centering Desirable Properties}}} & Accuracy & Low & High & High \\
 & Range & Low & Medium & High \\
 & Robustness to Impersonation & Low & Medium & High \\ 
 & Ease of Use & High & Medium & Low \\
 & Discretion & Low & Medium & High \\ \midrule
\multicolumn{2}{c}{Required Bystander Device} & None & LED Emitter & UWB Tag \\
\multicolumn{2}{c}{Cost} & Free & Cheap & Expensive \\
\bottomrule
\end{tabular}%
}
\vspace{-\baselineskip}
\end{table}

%% file: sections/motivation.tex
\section{Problem Setup and Terminology}\label{sec:motivation}

The proliferation of camera-enabled devices makes it so that one cannot currently expect privacy in shared and public spaces. People use smartphones  to capture moments all the time. Smart glasses (e.g., RayBan Meta Glasses~\cite{RayBanMeta2023}) only exacerbate this trend with their ambient recording and AI-capabilities to process the recorded images and videos. 

We consider the {\em bystander privacy problem} depicted on Figure~\ref{fig:overview}.
A \textit{user} is the individual operating a camera-enabled device. A \textit{bystander} is any non-user within the device's field of view who has not explicitly consented to data capture. There maybe one more bystanders in the field of view of the camera. The problem arises when the camera on a user's device captures data that can record, and identify a bystander or infer their sensitive attributes, creating a conflict between the user's utility and the bystander's right to privacy~\cite{corbett2023securing}. In this paper, we focus specifically on protecting the privacy of bystanders' {\em face}, which is a sensitive biometric.

A bystander communicates their \textit{privacy preference}
({\em i.e.} whether they want their face to be blurred or unblurred), to the user's device through a \textit{signaling modality}, which is the specific communication channel used. The user's device, then implements the bystander's preference, by detecting their face on the video stream and blurring/unblurring it accordingly, in real-time. This is a strong privacy protection, as it is applied before the video stream is stored or transmitted outside the camera-enabled device.

We also consider a \textit{bystander impersonation attack}, where an adversary attempts to change the bystander's privacy preference. For example, an adversary could position themselves between the bystander and the camera and signal a preference. This problem can also arise from non-malicious users, who stand (and signal from locations) close to the bystander, {\em e.g.,} in crowded spaces). To counter both adversarial and accidental such actions, we perform \textit{signal-source validation} to correctly bind each signal to the face of the originating bystander, in ways that depend  on each modality. 

%% file: sections/related_work.tex
\section{Related Work} \label{sec:related}

\begin{figure*}[t]
  \centering
  \includegraphics[width=0.9\textwidth]{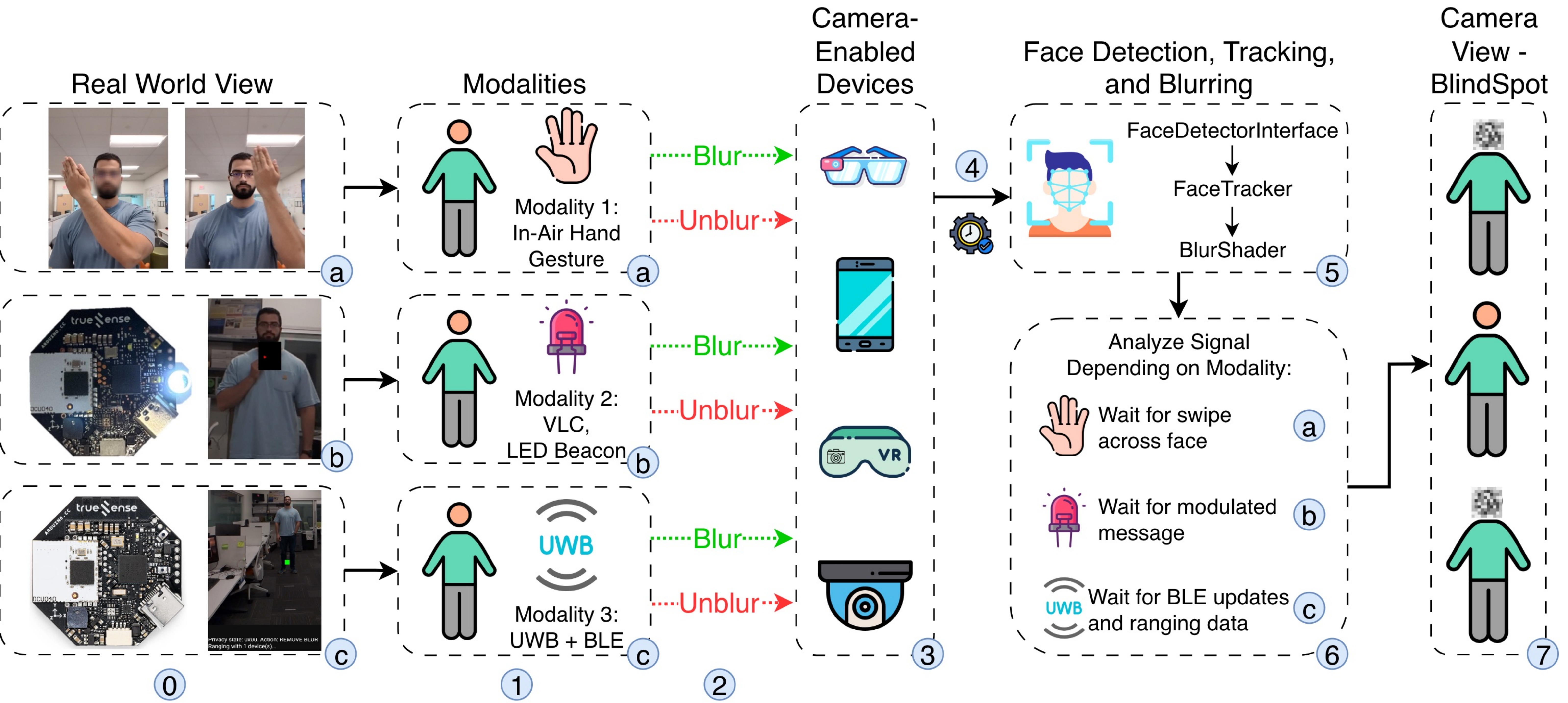}
  \captionsetup{font={footnotesize}}
  \caption{Overview of our system. (0) In the \textbf{Real World View}, bystanders can use different (1) modalities to signal their (2) privacy preference (e.g., 'Blur' or 'Unblur'). (0.a) Device-free hand gesture, (0.b) Arduino Stella w/ LED Diode in SWO pin, (0.c) Arduino Stella w/ UWB chip. They communicate this choice using one of three \textbf{Privacy Signaling Modalities}: (1.a) Modality 1 (Hand Gesture), (1.b) Modality 2 (VLC, LED Beacon), or (1.c) Modality 3 (UWB + BLE). (3) These signals are detected by nearby \textbf{Camera-Enabled Devices} (e.g., smart glasses, smartphones). (4) Frame processing is done in real-time. (5) Regardless of modality chosen, the devices perform \textbf{Face Detection, Tracking, and Blurring} to track all individuals in real-time. The detected faces data is passed to (6) a module that \textbf{Analyzes the Signal Depending on Modality}, which waits for a specific trigger. For example, (6.a) a hand swipe gesture, (6.b) a modulated LED message, or (6.c) a UWB/BLE data update. (7) This analysis determines the final action, resulting in the \textbf{Camera View - \tool{}}, where each individual's respective privacy preference (Blur or Unblur) is applied. This system empowers individuals with direct control over their visual privacy in ubiquitous sensing environments.}
  \label{fig:overview}
  %\vspace{-\baselineskip}
\end{figure*}

\subsection{Bystander Privacy Systems} The challenge of protecting bystander privacy in the presence of mobile and AR devices is well-documented. Corbett et al.~\cite{corbett2023securing} categorizes solutions into ``implicit'' and ``explicit'' systems, noting that an ideal real-time, on-device solution that balances usability, protection, and consent does not yet exist. Our work aims to fill this gap.

Implicit systems place control in the hands of the device user. For example, \textbf{BystandAR}~\cite{corbett2023bystandar} leverages the user's eye gaze and voice commands to infer which individuals are consenting vs. non-consenting bystanders. While effective, agency remains with the user of the device, not with the bystander. This leaves the bystander without a direct mechanism to assert their preferences, creating a need for a new architecture that grants them explicit agency over their own visual data. Our system fundamentally differs by providing direct, unambiguous control to the bystander.

Other systems are limited by their reliance on network connectivity or biometric identification. \textbf{I-Pic}~\cite{aditya2016ipic} offloads frames to a cloud service for anonymization, which introduces latency and fails in offline environments. \textbf{Cardea}~\cite{shu2018cardea} uses facial recognition to apply pre-registered privacy preferences, creating a centralized database of sensitive biometric data. In contrast, our entire pipeline is on-device, real-time, and biometric-free, avoiding both the latency of cloud processing and the privacy risks of facial recognition.

\subsection{Gestural and Tangible Interfaces} Our work translates user-centric design principles into a functional, bystander-centric system. Koelle et al.~\cite{koelle2018your} performed foundational user studies to identify socially acceptable gestures for privacy negotiation but did not implement a real-time system. Our velocity-based gesture recognizer is a similar implementation of these principles, designed for robustness on commodity mobile devices.

Prior work on gestural control has focused on user-centric interfaces, for instance, using head movements \textbf{GlassGesture}~\cite{yi2016glassgesture} or eye movements \cite{mohamed2025eye} for device control. Our first modality uses gestures for bystander-centric signaling.

The concept of ``Tangible Privacy,'' proposed by Ahmad et al.~\cite{ahmad2020tangible}, advocates for ``high-assurance'' physical controls like lens caps to bridge the gap between perceived and actual privacy. Our work evolves this concept from a static control on the device to a mobile, active token in the bystander's hand. Our VLC-based \seqsplit{Privacy Beacon} and our UWB tag are tangible mechanisms that provide the bystander with direct, high-assurance control over their privacy, even at a distance.

\subsection{Optical and RF Signaling}
\subsubsection{Optical Systems}
Prior systems have used optical markers for machine-to-machine tasks. \textbf{MarkIt}~\cite{raval2014markit} uses passive visual markers to protect static regions in a video feed, while \textbf{All that GLITTERs}~\cite{allthatglitters} uses active, blinking LEDs for anchoring AR content. Our VLC modality is a significant advancement of the VLC privacy signaling principles first introduced in our preliminary work \textbf{BystandARIA}~\cite{alaaraj2025bystandaria} and the foundational decoding logic from Salam and Al Aaraj~\cite{aaraj2024decoding}. While these earlier systems established the viability of using modulated LEDs for privacy signaling, our current work introduces a far more robust, Viterbi-based decoding algorithm. This new approach moves beyond simple frame differencing and contour analysis, providing resilience to movement and enabling the tracking of multiple, simultaneous signals in complex, dynamic environments. While specialized hardware like the event camera in \textbf{Helios 2.0}~\cite{bhattacharyya2025helios} enables ultra-low-power sensing, our system is designed for commodity camera-enabled devices to prioritize accessibility.

\subsubsection{RF Systems} 
Ultra-Wideband (UWB) is a radio technology that enables high-precision localization through short-pulse transmissions. This allows for accurate distance estimation via time-of-flight (ToF) measurements, a process known as \textbf{ranging}. Modern UWB antenna arrays also compute angle-of-arrival (AoA) data from the signal's phase differences, yielding the precise direction of a tag in addition to its distance. Ranging data includes distance, azimuth, and elevation. While UWB is well-established for secure ranging in applications like digital keys, its use for real-time spatial localization of bystanders against a live video feed is novel. Our work is the first to use UWB's angle-of-arrival data to create a link between the signal and a specific individual's face on a screen, enabling a new form of validated privacy signal control that is robust to visual occlusion and spoofing.

%% file: sections/sys_design.tex
\section{System Design}
\label{sec:design}

Our system provides real-time, on-device bystander privacy management through three distinct signaling modalities. The architecture comprises of (1) a common processing pipeline for face detection, tracking, and blurring on the user's device, and (2) signaling mechanisms between the bystander and the user to signal privacy preferences using three different modalities. This design achieves all desired features outlined in our introduction: it is biometric-free, enables direct bystander control, operates locally in real-time, and uses reliable signaling mechanisms. The overview of the system is shown in Figure~\ref{fig:overview}. \tool{} is implemented on a Google Pixel 8 Pro that is processing a 1280x960 stream at 30 FPS.

A core novel design aspect of \tool{} is using the on-screen face bounding box as a common reference point for signal-source validation. The system uses the position and size of this box to bind and validate the signal's origin to the corresponding bystander sending it. This check is done for all three modalities. This makes the systems robust at rejecting false positives, whether from accidental triggers or adversarial attacks, as detailed in the following subsections.

\subsection{Face Detection, Tracking, and Blurring}\label{subsec:face_tracking}
We refer as ``face pipeline'' to the component that performs real-time face detection, tracking, and blurring (Step 5, Figure~\ref{fig:overview}). This pipeline uses established techniques and is a necessary component of the \tool{} system, although not our core contribution. We implemented it as a single modality-agnostic backend to simplify the architecture and provide a common data source for all signaling methods. 

\noindent\textbf{Selection of Processing Resolution.}
We select a processing resolution of 1280x960; details provided in Appendix~\ref{app:proc_resolution}.

\noindent\textbf{Face Detection.}
The face detection is abstracted behind a \texttt{\seqsplit{FaceDetectorInterface}} to enable selection of the fastest and most accurate inference backend. We used this interface to evaluate three candidates: the MediaPipe Face Detector, and a YOLOv12n-face model using both the TFLite and NCNN runtimes. Based on its superior inference speed, which was critical for our real-time target, we chose the NCNN runtime.

\noindent\textbf{Distance Estimation.} We estimate the distance of the bystander based on the size of the face bounding box using an inverse projection model. See details in Section~\ref{subsubsec:distance-face}.

\noindent\textbf{Face Tracking.}
Stateless, per-frame detections are fed into our \texttt{\seqsplit{FaceTracker}} component, which assigns a persistent integer ID to each individual to maintain temporal identity across frames. The tracker implements a spatial proximity algorithm, associating new detections with existing tracks by calculating the Euclidean distance between bounding box centroids. A match is confirmed if this distance falls below a threshold of 0.25 in normalized screen coordinates. To ensure robustness against transient occlusions, the \texttt{\seqsplit{FaceTracker}} maintains an ID for a face for up to 1000\,ms after it is last seen before pruning the track.

\noindent\textbf{Face Blurring.}
For each frame, a custom OpenGL renderer (\texttt{\seqsplit{BlurShader}}) applies a shader to the bounding box of any face ID flagged for anonymization. This operation is executed on the GPU, ensuring anonymization occurs with minimal latency. We use a multi-pass Gaussian blur as a proof-of-concept. The rendering pipeline is modular and can readily accommodate stronger, non-reversible anonymization techniques, such as pixelation or full redaction.

\subsection{Signaling Modalities}
A core contribution of this work is the introduction of three novel signaling modalities that allow a bystander to communicate privacy preferences to the user's device. These methods are summarized as Steps 1 and 6 in Figure~\ref{fig:overview}.

\subsubsection{Modality 1: Hand Gesture}
\begin{figure}[t]
  \centering
  \includegraphics[width=\columnwidth]{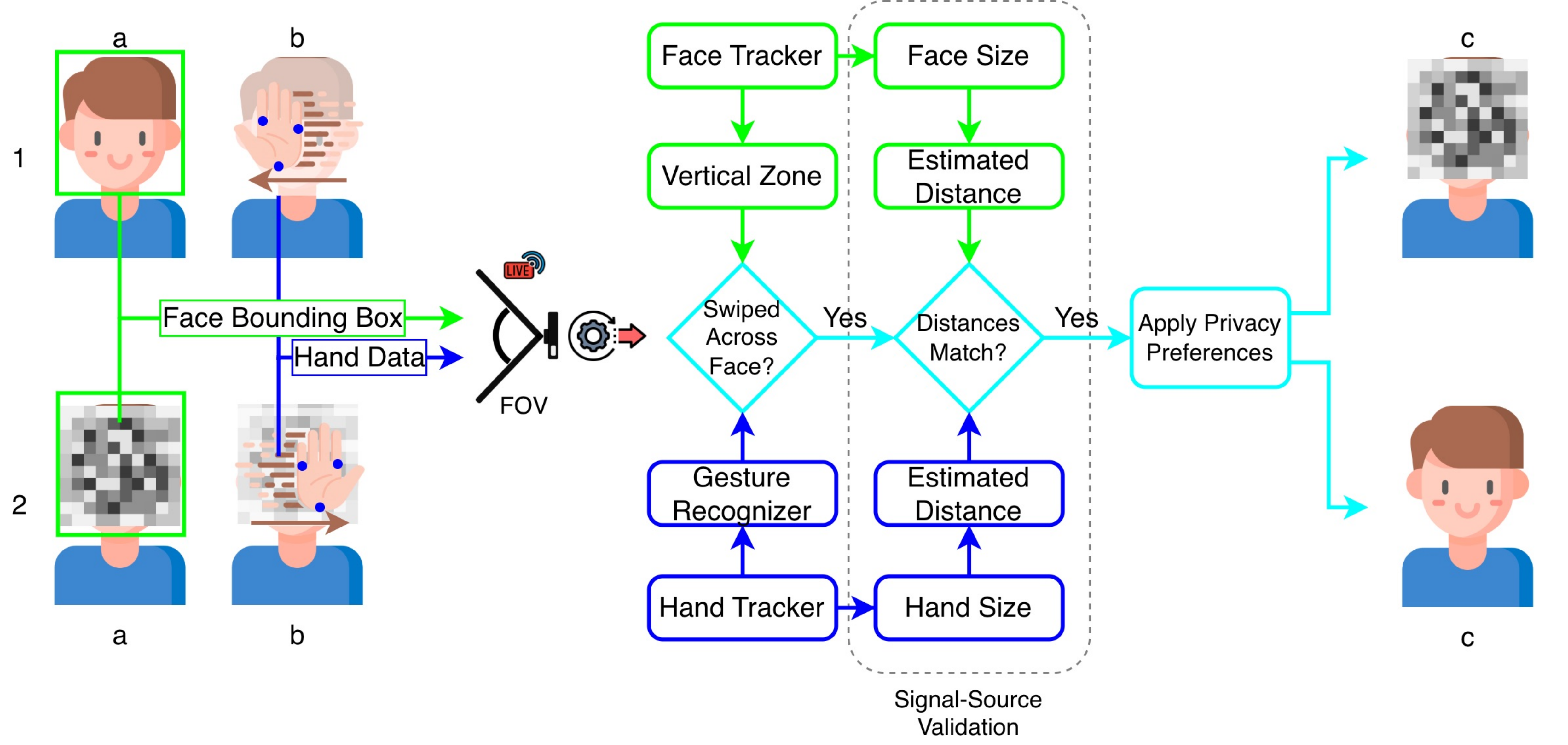}
  \captionsetup{font={footnotesize}}
  \caption{Illustration of the ``Hand Gesture'' signaling modality. \textbf{Top (1):} (a) A bystander's initial state. (b) The bystander performs a `blur' gesture, swiping from their left to right, which is captured and processed by a camera's Field of View (FOV). (c) The system then blurs the bystander's face in the camera's view. \textbf{Bottom (2):} (a) The bystander's face is initially blurred. (b) The bystander performs an `unblur' gesture, swiping from their right to left. (c) The system removes the blur.}
  \label{fig:gesture}
  \vspace{-\baselineskip}
\end{figure}

The first modality operates over a kinematic channel, using a custom gestural protocol to interpret a swipe gesture across the face. We selected this protocol for its high accessibility, as it is intuitive and requires no special hardware from the bystander. The overview of this modality is shown in Figure~\ref{fig:gesture}.

\noindent\textbf{Hand Detection Algorithm.} We use the MediaPipe HandLandMarker model. See details in Appendix~\ref{app:hand-model}.

\noindent\textbf{Hand Detection and Tracking.}
To recognize a dynamic gesture, the system must first establish a stable identity for a hand across multiple frames. The pipeline begins by processing the video stream with the MediaPipe Hand Landmarker~\cite{mediapipe_hand}, which provides the 21 normalized 3D landmarks for each detected hand. These raw detections are then passed to our \texttt{\seqsplit{HandTracker}} component, which assigns a persistent integer ID to each hand by implementing a proximity-based matching algorithm. For each existing track, it calculates the Euclidean distance between its wrist landmark from the previous frame and the wrist landmarks of all new detections in the current frame. A match is confirmed if the distance is below an empirically determined threshold (\texttt{\seqsplit{maxDistance}}). New IDs are assigned to any unmatched detection.

\noindent\textbf{Velocity-Based Gesture Recognition.}
To ensure robustness against variable frame rates and stuttered motion, we analyze the hand's velocity rather than relying on simple positional checks. An approach based on hand-face bounding box intersection is naive. Our \texttt{\seqsplit{GestureRecognizer}} instead implements a velocity-based approach that analyzes the dynamics of the hand's movement. For each tracked hand, the recognizer maintains a temporal history of its palm center's horizontal position over the last 300\,ms (\texttt{\seqsplit{HISTORY\_MAX\_AGE\_MS}}). The palm center is calculated as the average position of the wrist, index finger MCP, and pinky MCP landmarks to provide a stable anchor point.

A swipe is registered only when the horizontal displacement between the hand's current position and \textit{any} point within its recent 300\,ms history exceeds a dynamic threshold. This threshold is defined as a fraction (\texttt{\seqsplit{MIN\_SWIPE\_DISTANCE\_FACTOR}}) of the width of the associated face's bounding box. This scaling ensures that the required gesture magnitude adapts to the bystander's distance from the camera. A user further away (with a smaller face bounding box) can use a physically smaller swipe, making the interaction feel consistent regardless of distance.

\noindent\textbf{Directional Control and Face Association.}
After recognizing a swipe, the system must map the gesture to a specific privacy command and attribute it to the correct individual. The system uses the direction of the gesture's displacement vector to determine the privacy signal intended. A swipe from the user's left to their right (a positive horizontal displacement) signals a request for anonymization. Conversely, a swipe from right to left revokes this request. The hand is associated with a face by finding the \texttt{\seqsplit{TrackedFace}} whose bounding box centroid has the minimum Euclidean distance to the hand's wrist landmark.

\noindent\textbf{Signal-Source Validation.}
To minimize false positives from incidental hand movements and adversarial threats, the gesture recognizer incorporates a series of strict constraints.
\begin{enumerate}[nosep, leftmargin=*]
    \item \textbf{Vertical Constraint:} A gesture is considered valid only if the hand's palm center remains within a constrained vertical zone during the swipe. This zone extends a fraction (\texttt{\seqsplit{VERTICAL\_TOLERANCE\_FACTOR}}) of the face's height above and below the face's bounding box. This check effectively rejects signals from hands moving elsewhere in the frame, such as a user waving to someone else.
    \item \textbf{Dual Cooldown Mechanism:} After a successful gesture is registered, the system applies a 1500\,ms cooldown to both the hand's ID and the associated face's ID. The hand cooldown prevents a single, continuous hand motion from being erroneously registered as multiple distinct gestures. The face cooldown prevents a recently-toggled privacy state from being immediately toggled back, for instance, by an accidental gesture from the user's other hand.
    \item \textbf{Geometric Consistency:} To prevent bystander impersonation attacks from a adversarial actor's hand, the system validates that the hand and face are at a consistent distance. It accepts a gesture only if the distance inferred from the hand's on-screen size matches the distance inferred from the face's bounding box (10\% tolerance), as detailed in our evaluation in Section~\ref{sec:eval-parameters}.
\end{enumerate}
Together, these constraints, analyzing not just position but also velocity, and enforcing strict spatial and temporal rules, create a robust gesture detection system resilient to accidental activation without requiring any explicit hardware.

\begin{figure}[t]
  \centering
  \includegraphics[width=1\columnwidth]{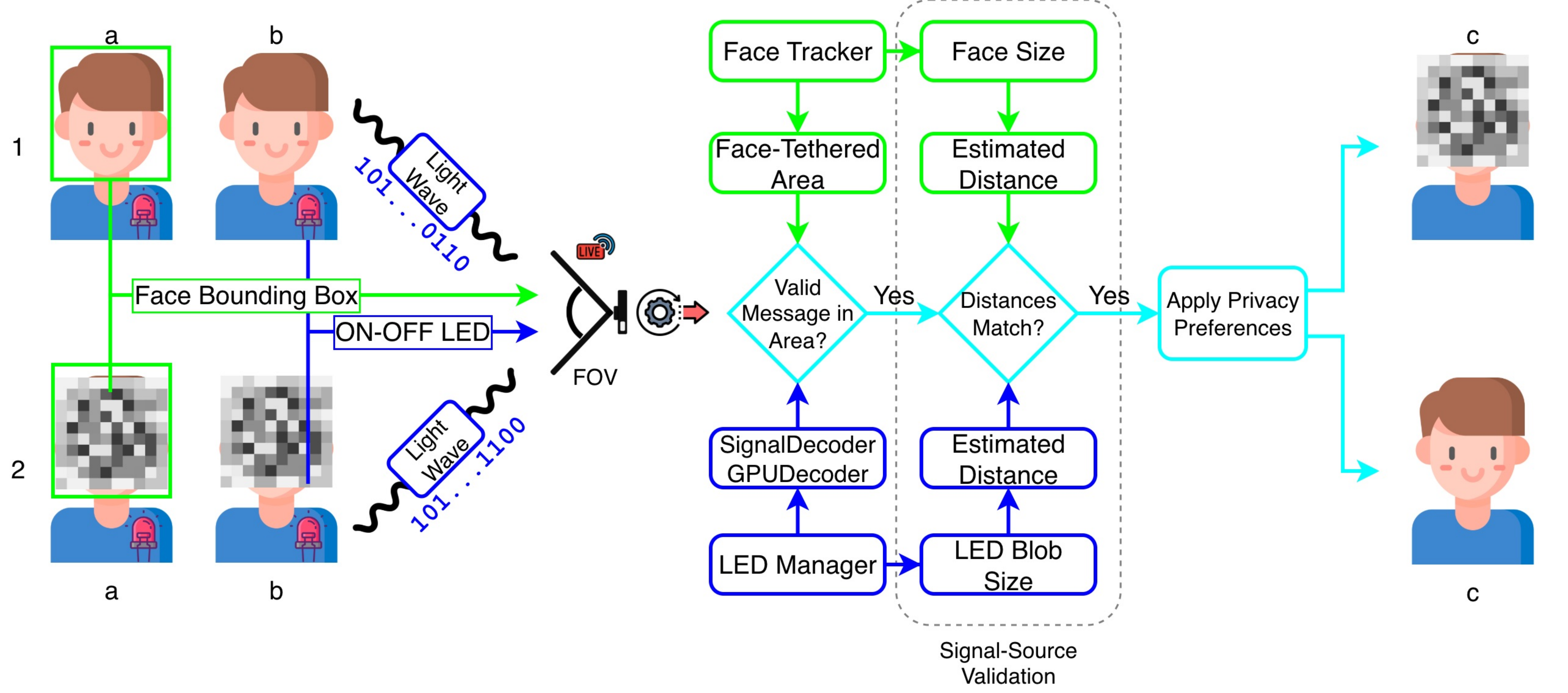}
  \captionsetup{font={footnotesize}}
  \caption{Illustration of the privacy preference signaling process using visible light communication (VLC). \textbf{Top (1):} (a) A bystander's initial unblurred state. (b) The bystander's device (represented by the LED) modulates a `blur' message into a light wave, transmitted by rapidly switching the LED ON/OFF according to a predefined bit sequence (e.g., preamble, body, CRC), which is then captured and processed by a camera's Field of View (FOV). (c) The system applies the `blur' preference to the bystander's face. \textbf{Bottom (2):} (a) The bystander's face is initially blurred. (b) The device transmits an `unblur' message via VLC. (c) The system removes the blur.}
  \label{fig:led}
  \vspace{-\baselineskip}
\end{figure}

\subsubsection{Modality 2: Visual Light Communication}
The second modality uses an optical channel, enabling signaling via a custom On-Off Keying (OOK) protocol for Visual Light Communication (VLC). We chose OOK for its simplicity and low power consumption, which is well-suited for transmitting small command packets from a battery-powered device. The device transmits a command by modulating its LED, and the phone's camera decodes the signal. To overcome the prohibitive computational cost of full-frame VLC decoding, our design introduces a \textit{face-attached search} strategy. This approach uses the detected face's bounding box to define a small, targeted search area for the optical signal, drastically reducing processing overhead. The overview of this modality is shown in Figure~\ref{fig:led}. Camera configuration setting details are found in Appendix~\ref{app:camera-config}.

\noindent\textbf{Search Areas and GPU-Accelerated Processing.}
To avoid the cost of full-frame analysis, our face-attached search strategy leverages the GPU to process only small, relevant regions of the image. For each tracked face, the system dynamically defines a rectangular search area below its bounding box, based on the assumption that a bystander will wear the beacon on their torso (scalable to other parts of the upper body). A custom fragment shader (\texttt{\seqsplit{GPUDetector}}) analyzes only pixels within these search areas. Furthermore, the shader uses the distance-to-blob-size relationship characterized in our evaluation (Appendix~\ref{app:led-size}) to dynamically set the kernel size for finding local maxima, ensuring efficient and accurate LED detection across different distances. Immediately following the render pass, we use \texttt{\seqsplit{glReadPixels}} on these small areas. This targeted readback minimizes the GPU-to-CPU data transfer bottleneck, a common performance limiter.

\noindent\textbf{Multi-Path Viterbi-based Decoding.}
The Viterbi algorithm is a dynamic programming method for finding the most likely sequence of hidden states. We adapt this approach to decode the optical signal because it is inherently robust to the challenges of a mobile environment, such as intermittent visibility from motion blur and temporary occlusions. Instead of making a hard decision on each frame, our Viterbi-like multi-path tracker maintains a set of the most probable bit sequences (\textit{paths}), allowing it to recover the correct message even with several corrupted or missing frames. The process unfolds in three stages per frame:
\begin{enumerate}[nosep, leftmargin=*]
    \item \textbf{Blob Detection:} For each search area's pixel buffer, a Breadth-First Search (BFS) identifies all distinct groups of highlighted pixels, or \textit{blobs}.
    \item \textbf{Path Extension:} The algorithm extends all active paths from the previous frame by associating each with the spatially nearest available blob, appending a `1' bit. If no blob is found, it appends a `0' bit and adds a cost penalty. New paths are initiated for any unassigned blobs.
    \item \textbf{Pruning:} After extension, the list is sorted by cost and pruned to a fixed number of the most likely candidates (\texttt{\seqsplit{MAX\_ACTIVE\_PATHS}}).
\end{enumerate}

\noindent\textbf{Message and Signal-Source Validation.}
To finalize the decoding and validate the signal's authenticity, the system must confirm both the packet's integrity and its geometric origin. When a path reaches the required 18-bit length, a \texttt{\seqsplit{SignalDecoder}} validates its preamble and checksum. If the packet is valid and contains a recognized command, the system performs a final geometric validation check. It compares the distance (10\% tolerance) inferred from the LED blob's on-screen area against the distance inferred from the associated face's bounding box. A signal is accepted only if these two distance estimates are consistent.

\subsubsection{Modality 3: UWB Spatial Localization}
\begin{figure}[t]
  \centering
  \includegraphics[width=1\columnwidth]{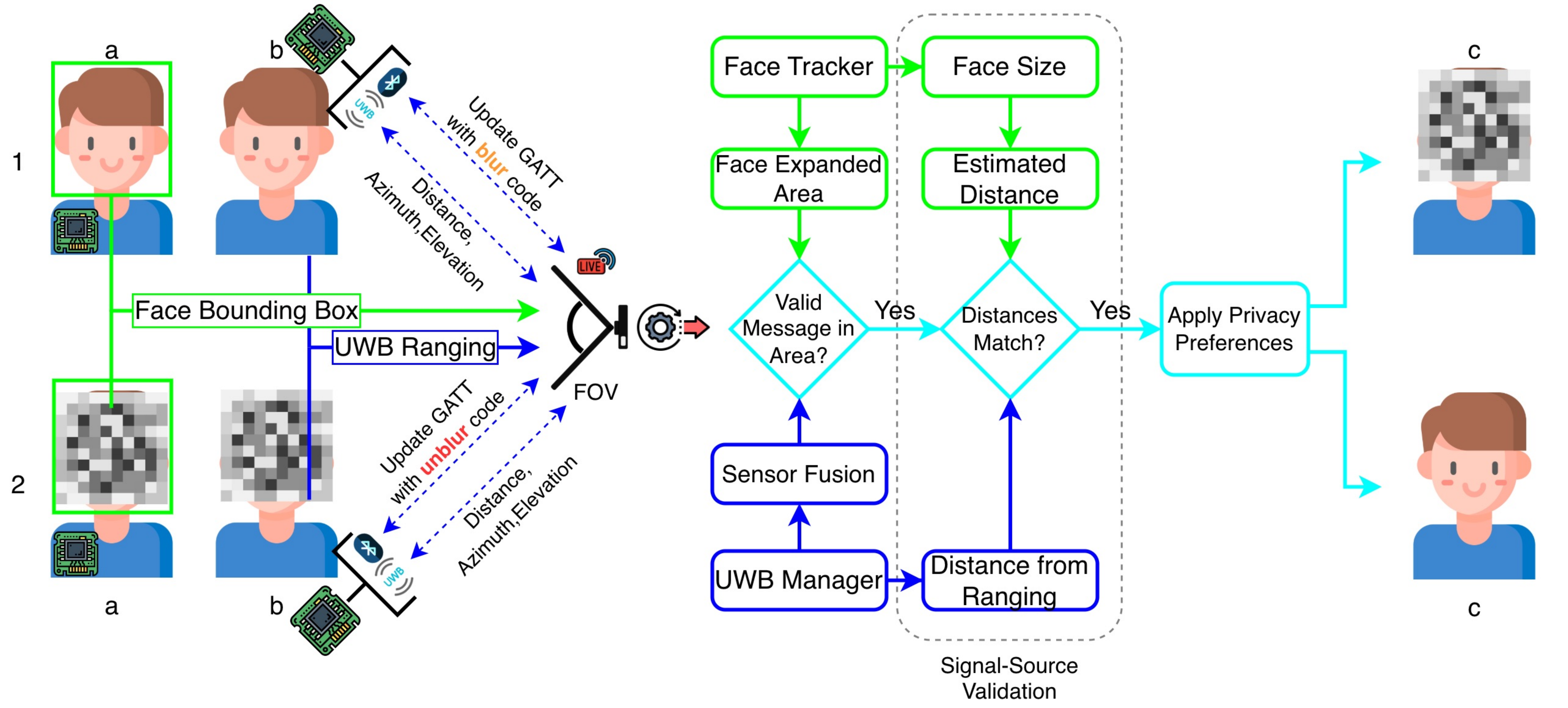}
  \captionsetup{font={footnotesize}}
  \caption{Illustration of the privacy signaling process using Ultra-Wideband (UWB) and Bluetooth. \textbf{Top (1):} (a) A bystander's initial unblurred state. (b) The bystander sends a `blur' command thorugh the UWB tag and then transmits its precise location (Distance, Azimuth, Elevation) to the camera-enabled device (e.g., ``Update GATT with blur code''). (c) The system processes this information and applies a blur filter to the bystander's face. \textbf{Bottom (2):} (a) The bystander's face is initially blurred. (b) The device transmits its location and sends an `unblur' command. (c) The system removes the blur.}
  \label{fig:uwb}
  \vspace{-\baselineskip}
\end{figure}

The third modality provides the strongest signal-source validation using Ultra-Wideband (UWB), a radio technology designed for precise indoor localization and now integrated into modern smartphones. While UWB's time-of-flight ranging offers inherent robustness against impersonation attacks, its power consumption is prohibitive for continuous operation. To address this, we designed a hybrid BLE+UWB protocol that uses UWB only for on-demand spatial localization. This approach minimizes power while retaining the benefits of the centimeter-level accuracy of UWB. Our prototype uses a custom tag, but the protocol is designed for future deployment directly on UWB-enabled phones. The overview of this modality is shown in Figure~\ref{fig:uwb}.

\begin{figure}[t]
  \centering
  \includegraphics[width=1\columnwidth]{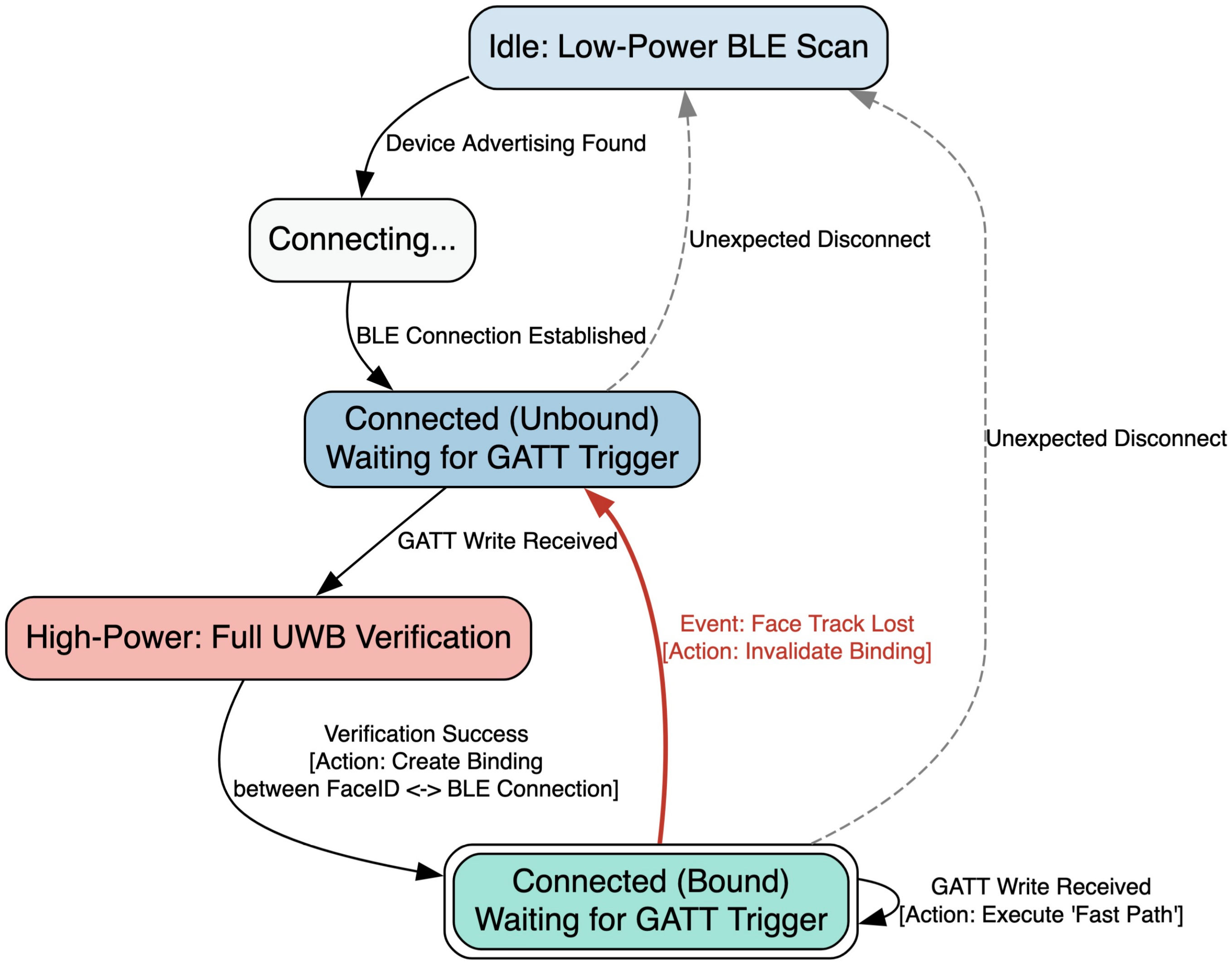}
  \captionsetup{font={footnotesize}}
  \caption{State machine for the hybrid UWB+BLE protocol's power management. The system distinguishes between a \textbf{\texttt{Disconnected (Unbound)}} and a \textbf{\texttt{Connected (Bound)}} state. A device is initially \texttt{Unbound}, where a GATT trigger forces a \texttt{Full UWB Verification}. A successful verification creates a binding between the device's MAC address and a \texttt{FaceID}, transitioning the system to the \texttt{Bound} state. Subsequent GATT triggers in this state use the low-power \texttt{Fast Path}. Crucially, the system reverts to the \texttt{Unbound} state via the \textbf{Face Track Lost} event. This transition invalidates the binding the moment the associated face is no longer visible, ensuring stale associations are purged and forcing re-verification for any future interaction.}
  \label{fig:uwb-ble-state}
  %\vspace{-\baselineskip}
\end{figure}

\noindent\textbf{Hybrid BLE+UWB Protocol.} To minimize power consumption, our protocol uses low-power BLE for persistent connection and initial privacy preference signaling, reserving the power-intensive UWB radio for explicit localization events. The tag advertises a BLE service, and a privacy request is sent via a GATT characteristic write. This notification acts as a trigger for the \tool{}'s \texttt{\seqsplit{UwbManager}}. The manager then decides whether to initiate a full UWB spatial localization or, if the tag is already trusted, use a power-saving cache mechanism we call the ``Fast Path''. The state machine for this protocol is shown in Figure~\ref{fig:uwb-ble-state}.

\noindent\textbf{Full Spatial Localization Path.}
When a request is received from an unbound (Face $\leftrightarrow$ BLE Connection) tag, the system initiates this path to establish a secure spatial binding. The goal is to map the tag's physical 3D position to a 2D screen coordinate and link it to the face of the signaling bystander.

\begin{enumerate}[nosep, leftmargin=*]
    \item 
    \textbf{Burst Ranging:} The manager commands the tag to begin a UWB ranging session and collects a burst of \texttt{\seqsplit{TOTAL\_READING\_COUNT}} (e.g., 15) angle-of-arrival readings.
    \item 
    \textbf{Data Smoothing:} To produce a stable position estimate free from radio interference and transient fluctuations, the system averages the final \texttt{\seqsplit{READINGS\_TO\_AVERAGE}} (e.g., 3) measurements from the burst. This yields a stable estimate of the tag's azimuth and elevation.
    \item 
    \textbf{Geometric Projection:} The smoothed ranging data (distance, azimuth, elevation) are passed to a projection module (\texttt{\seqsplit{SensorFusion}}). This module first checks if the tag's angles fall within the camera's known Field of View (FoV). If so, it performs a linear mapping of the tag's angular position to a 2D screen coordinate.
    \item 
    \textbf{Binding:} Successful binding occurs if the projected 2D coordinate falls within the expanded bounding box of a face currently tracked by the face tracking module (see Section~\ref{subsec:face_tracking}). The system then associates the tag and the bystander's face. The width of the bounding box is expanded by a factor of 2.5 (Appendix~\ref{app:uwb-face-factor}).
\end{enumerate}

\noindent\textbf{Power Management.}
To eliminate the power, latency cost, and single connection capability of UWB ranging for subsequent requests from a trusted tag, the ``Fast Path'' serves as a temporary authorization cache. After a successful spatial binding, the system creates a trusted link in a hash map, mapping the tag's unique BLE MAC address to the verified \texttt{\seqsplit{FaceID}}. For all subsequent BLE triggers from that MAC address, the system checks if the linked face is still visible. If it is, the privacy state is toggled immediately, bypassing the UWB ranging session entirely.

\noindent\textbf{Signal-Source Validation.}
UWB's ranging provides a strong baseline against attacks. Our protocol builds on this with three additional layers of validation. First, a tag is only trusted after an initial spatial binding verifies its projected screen coordinate corresponds to a bystander's face. Second, we check if the measured UWB distance is almost the same size as the matched face's bounding box (10\% tolerance). A significant discrepancy indicates an anomaly and causes the validation to fail. Finally, the ``Fast Path'' cache is secured by invalidating any trusted link the moment its associated face is no longer visible, preventing an adversary from exploiting a stale association.

%% file: sections/evaluation.tex
\section{Evaluation}
\label{sec:evaluation}

We conduct a comprehensive evaluation to demonstrate the feasibility of \tool{} as a real-time, on-device, bystander-controlled privacy management system and to analyze the trade-offs between its three signaling modalities. Our experiments are designed to answer three key questions: (1) How effective and robust is each signaling modality against real-world challenges such as varying distance, user motion, and ambient lighting? (2) How does the system's performance scale with an increasing number of bystanders? (3) What is the  computational overhead of the complete system, and does it meet the 30 FPS target on a commodity mobile device?

We first validate key design parameters used across the modalities. We then present an end-to-end performance analysis for each signaling method. In addition, we compare our results to the state-of-the-art bystander privacy systems.

\subsection{Model and Parameter Selection} \label{sec:eval-parameters}
Next, we justify our choice of parameters and adaptive logic detailed in our system design (Section~\ref{sec:design}). These results demonstrate how \tool{} is optimized for robust performance across a range of conditions.

\begin{figure}[t]
    \centering
    \includegraphics[width=1\columnwidth]{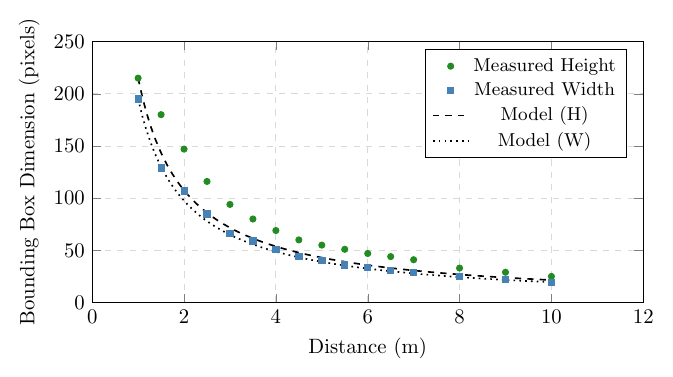}
    \captionsetup{font={footnotesize}}
    \caption{Face bounding box dimensions as a function of distance. Both measured height and width closely follow the theoretical inverse projection model.}
    \label{fig:distance_characterization}
    \vspace{-\baselineskip}
\end{figure}

\subsubsection{Face Detection: Distance Estimation}\label{subsubsec:distance-face}
\tool{} uses on-screen face size as a primary input for its signal-source validation. To verify this approach, we first characterized the relationship between physical distance and the pixel height and width of a subject's face bounding box. Figure~\ref{fig:distance_characterization} plots our empirical measurements against a theoretical inverse projection model ($size \propto 1/distance$). The measured data closely tracks the model, confirming that bounding box height and width are a reliable proxy for distance. At \texttt{1}\,m, the average face height was \texttt{215}\,pixels, decreasing to \texttt{23}\,pixels at our maximum test distance of \texttt{10}\,m. \tool{} leverages this validated geometric relationship to cross-check the hand size of gesture, blob size of LED, and ranging data (distance) from the UWB modality.

\subsubsection{Modality 1: Hand Distance Estimation}\label{subsec:eval_hand}
We do the same evaluation of distance estimation based on hand size as Section~\ref{subsubsec:distance-face}. Check Appendix~\ref{app:hand-size} for details.

\subsubsection{Modality 2: VLC Design and Validation}
The VLC modality design balances decoding reliability against motion robustness and incorporates a geometric check for validation.

\begin{figure}[t]
    \centering
    \includegraphics[width=1\columnwidth]{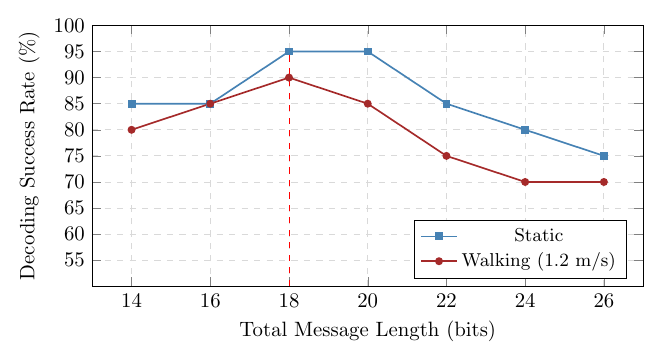}
    \captionsetup{font={footnotesize}}
    \caption{VLC decoding success rate vs. message length at 3\,m. The 18-bit format provides the best performance for walking users, balancing reliability and motion robustness.}
    \label{fig:vlc_tradeoff}
    \vspace{-\baselineskip}
\end{figure}

\noindent\textbf{Message Size Optimization.} The message length presents a fundamental trade-off: shorter messages have lower latency but are noise-sensitive, while longer messages are more reliable but vulnerable to motion blur. To quantify this, we evaluated the decoding success rate of seven different message formats (from 14 to 26 bits) at a 3\,m distance under both static and walking (\texttt{1.2}\,m/s) conditions. Each test was repeated 20 times. Figure~\ref{fig:vlc_tradeoff} shows that for walking users, the success rate peaks at 18 bits before degrading as longer transmission times increase the probability of motion-induced errors. Our chosen 18-bit format (6-bit preamble, 8-bit payload, 4-bit CRC) achieves a 90\% success rate under motion, balancing robustness and latency (\texttt{600}\,ms transmission time).

\noindent\textbf{LED Distance Estimation.} We do the same evaluation of distance estimation based on LED size as Section~\ref{subsubsec:distance-face}. Check Appendix~\ref{app:led-size} for details.

% Use table* for a two-column wide table. This will be very long.
\begin{table*}[t]
\centering
\small
\captionsetup{font=small}
\caption{End-to-end performance comparison of the three signaling modalities. We report Accuracy (Acc.), False Positive (FP) and False Negative (FN) rates, and end-to-end Latency in milliseconds. The UWB modality is unaffected by lighting conditions. As mentioned in Section~\ref{subsec:eval_hand}, hand tracking doesn't work past 3 meters.}
\label{tab:performance}
\begin{tabular}{@{}lccccccccccccccc@{}}
\toprule
& \multicolumn{5}{c}{\textbf{Modality 1: Hand Gesture}} & \multicolumn{5}{c}{\textbf{Modality 2: VLC Beacon}} & \multicolumn{5}{c}{\textbf{Modality 3: UWB Tag}} \\
\cmidrule(lr){2-6} \cmidrule(lr){7-11} \cmidrule(lr){12-16}
\textbf{Condition} & Acc. & FP & FN & Latency & & Acc. & FP & FN & Latency & & Acc. & FP & FN & Latency \\
& (\%) & (\%) & (\%) & (ms) & & (\%) & (\%) & (\%) & (ms) & & (\%) & (\%) & (\%) & (ms) \\
\midrule
\multicolumn{16}{l}{\textit{\textbf{Static User, Low Light}}} \\
\addlinespace[0.3em]
\hspace{1em}Near (3m) & 85.0 & 00.0 & 15.0 & 164 && 100.0 & 00.0 & 00.0 & 624 && \texttt{N/A} & \texttt{N/A} & \texttt{N/A} & \texttt{N/A} \\
\hspace{1em}Mid (6m) & \texttt{N/A} & \texttt{N/A} & \texttt{N/A} & \texttt{N/A} && 100.0 & 00.0 & 00.0 & 621 && \texttt{N/A} & \texttt{N/A} & \texttt{N/A} & N/A \\
\hspace{1em}Far (10m) & \texttt{N/A} & \texttt{N/A} & \texttt{N/A} & \texttt{N/A} && 90.0 & 00.0 & 10.0 & 619 && \texttt{N/A} & \texttt{N/A} & \texttt{N/A} & \texttt{N/A} \\
\midrule
\multicolumn{16}{l}{\textit{\textbf{Static User, Normal Light}}} \\
\addlinespace[0.3em]
\hspace{1em}1m & 100.0 & 00.0 & 00.0 & 150 && 100.0 & 00.0 & 00.0 & 632 && 85.0 & 00.0 & 15.0 & 1990 \\
\hspace{1em}2m & 100.0 & 00.0 & 00.0 & 166 && 95.0 & 00.0 & 5.0 & 621 && 90.0 & 00.0 & 10.0 & 1980 \\
\hspace{1em}3m & 90.0 & 5.0 & 5.0 & 200 && 95.0 & 00.0 & 5.0 & 624 && 95.0 & 00.0 & 5.0 & 1990 \\
\hspace{1em}4m & \texttt{N/A} & \texttt{N/A} & \texttt{N/A} & \texttt{N/A} && 95.0 & 00.0 & 5.0 & 635 && 95.0 & 00.0 & 5.0 & 1980 \\
\hspace{1em}5m & \texttt{N/A} & \texttt{N/A} & \texttt{N/A} & \texttt{N/A} && 95.0 & 00.0 & 5.0 & 631 && 95.0 & 00.0 & 5.0 & 1980 \\
\hspace{1em}6m & \texttt{N/A} & \texttt{N/A} & \texttt{N/A} & \texttt{N/A} && 95.0 & 00.0 & 5.0 & 642 && 100.0 & 00.0 & 00.0 & 1980 \\
\hspace{1em}7m & \texttt{N/A} & \texttt{N/A} & \texttt{N/A} & \texttt{N/A} && 95.0 & 00.0 & 5.0 & 608 && 100.0 & 00.0 & 00.0 & 1980 \\
\hspace{1em}8m & \texttt{N/A} & \texttt{N/A} & \texttt{N/A} & \texttt{N/A} && 95.0 & 00.0 & 5.0 & 655 && 100.0 & 00.0 & 00.0 & 1980 \\
\hspace{1em}9m & \texttt{N/A} & \texttt{N/A} & \texttt{N/A} & \texttt{N/A} && 90.0 & 00.0 & 10.0 & 619 && 100.0 & 00.0 & 00.0 & 1980 \\
\hspace{1em}10m & \texttt{N/A} & \texttt{N/A} & \texttt{N/A} & \texttt{N/A} && 85.0 & 00.0 & 15.0 & 637 && 95.0 & 00.0 & 5.0 & 1980 \\
\midrule
\multicolumn{16}{l}{\textit{\textbf{Static User, Bright Light}}} \\
\addlinespace[0.3em]
\hspace{1em}Near (3m) & 90.0 & 5.0 & 5.0 & 194 && 85.0 & 00.0 & 15.0 & 634 && \texttt{N/A} & \texttt{N/A} & \texttt{N/A} & \texttt{N/A} \\
\hspace{1em}Mid (6m) & \texttt{N/A} & \texttt{N/A} & \texttt{N/A} & \texttt{N/A} && 80.0 & 00.0 & 20.0 & 653 && \texttt{N/A} & \texttt{N/A} & \texttt{N/A} & \texttt{N/A} \\
\hspace{1em}Far (10m) & \texttt{N/A} & \texttt{N/A} & \texttt{N/A} & \texttt{N/A} && 65.0 & 00.0 & 35.0 & 667 && \texttt{N/A} & \texttt{N/A} & \texttt{N/A} & \texttt{N/A} \\
\midrule
\multicolumn{16}{l}{\textit{\textbf{Walking User, Normal Light}}} \\
\addlinespace[0.3em]
\hspace{1em}Near (3m) & 90.0 & 00.0 & 10.0 & 205 && 90.0 & 00.0 & 10.0 & 683 && 95.0 & 00.0 & 5.0 & 1980 \\
\hspace{1em}Mid (6m) & \texttt{N/A} & \texttt{N/A} & \texttt{N/A} & \texttt{N/A} && 85.0 & 00.0 & 15.0 & 684 && 100.0 & 00.0 & 00.0 & 1980 \\ 
\hspace{1em}Far (10m) & \texttt{N/A} & \texttt{N/A} & \texttt{N/A} & \texttt{N/A} && 80.0 & 00.0 & 20.0 & 693 && 95.0 & 00.0 & 5.0 & 1980 \\
\midrule
\multicolumn{16}{l}{\textit{\textbf{Multiple Users Static (6), Normal Light}}} \\
\addlinespace[0.3em]
\hspace{1em}Seated at Different Distances (2-6m) & 87.5 & 4.17 & 8.33 & 265 && 90.83 & 00.0 & 9.17 & 654 && 91.67 & 00.0 & 8.33 & 2009 \\
\midrule
\multicolumn{16}{l}{\textit{\textbf{Multiple Users Moving (6), Normal Light}}} \\
\addlinespace[0.3em]
\hspace{1em}Different Distances (2-6m) & 82.5 & 9.17 & 8.33 & 281 && 81.67 & 00.0 & 18.33 & 754 && 90.0 & 00.0 & 10.0 & 2043 \\
\bottomrule
\end{tabular}
\vspace{-1em}
\end{table*}

\subsection{Detection Performance}
\label{sec:eval-performance}

We evaluate the performance of each signaling modality under a range of controlled conditions. Our evaluation measures four key metrics: accuracy (a correct state toggle after a valid signal), false positive rate (an incorrect toggle without a signal), false negative rate (a failure to toggle after a valid signal), and end-to-end latency (time from signal initiation to on-screen blurring).

\subsubsection{Defining the Experimental Distance}
We selected our experimental distance range of \texttt{1}m to \texttt{10}m to correspond with the operational limits of the underlying face detection model. The feasibility of the system is fundamentally bound by the number of pixels required for reliable face detection. Our YOLOv12-based detector requires a face to be approximately \texttt{20x20} pixels, which is comparable to other modern detectors~\cite{mtcnn}. At our system's chosen 1280x960 processing resolution, a face at \texttt{10}m occupies a bounding box of roughly \texttt{20x25} pixels. This makes \texttt{10}m the effective maximum range for our experiments, allowing us to test the signaling modalities at the boundary of the vision pipeline's capability. The \texttt{1}m minimum distance represents a typical close-range interaction. This ensures our evaluation focuses on the performance of the signaling modalities themselves within a validated operational range.

\subsubsection{Experimental Setup}
To simulate real-world scenarios, we systematically varied four experimental parameters. For each condition, participants were given specific instructions to ensure consistent and repeatable trials.

    \noindent\textbf{Distance:} We positioned bystanders at 1-meter increments from the camera, ranging from 1\,m to 10\,m, to cover close-range, mid-range, and far-range interactions.

    \noindent\textbf{Lighting:} We tested under three lighting levels measured at the subject's position: dim indoor (100 lux), normal indoor (300 lux), and outdoors on a sunny day (1000 lux) to simulate various light ambient conditions.

    \noindent\textbf{Bystander Motion:} To capture both stationary and transient bystanders, we tested two motion levels. \textbf{(1) Static:} Bystanders were instructed to stand at a fixed position facing the camera and perform the signal. They were instructed to interact with each other normally to simulate a normal environment. \textbf{(2) Walking:} Bystanders were instructed to walk at a regular pace (around \textasciitilde1.2 m/s) across the camera's FOV while attempting to send the signal.

    \noindent\textbf{Bystander Count:} To assess performance in sparse and crowded scenes, we tested two configurations.\textbf{ (1) Single Bystander:} A single bystander was in the camera's FOV and performed the signaling task. \textbf{(2) Multiple Bystanders:} \textbf{Six} bystanders were present, instructed to signal at will while interacting normally. The N=6 configuration represents a high-load scenario designed to stress the system's real-time throughput, approaching the performance limits identified in our overhead analysis (Section~\ref{sec:eval-overhead}).

Each unique combination of conditions was tested 20 times per bystander (signaling user). For the multiple-user scenario, this resulted in 120 total trials. The aggregated results are summarized in Table~\ref{tab:performance}.

\subsection{System Overhead}
\label{sec:eval-overhead}

\subsubsection{CPU and GPU Overhead}
A key design goal of \tool{} is to perform all processing on-device while maintaining a real-time 30 FPS user experience. We quantify the computational cost and scalability of our system on a Google Pixel 8 Pro by measuring its resource utilization as the number of bystanders (N) increases from 1 to 10. Our architecture offloads the heavy vision processing to a background thread. Consequently, the primary performance bottleneck is the processing time of this background pipeline. We measure utilization as the ratio of the average processing time per frame to the target frame budget (33.3\,ms for 30 FPS; direct GPU usage percentage is inaccessible on non-rooted Android devices). A value over 100\% indicates that the system cannot process frames as fast as they arrive, leading to increased latency and eventual frame drops.

\begin{figure}[t]
\centering

% Subfigure for CPU Utilization
\begin{subfigure}{\columnwidth}
    \centering
    \includegraphics[width=0.8\columnwidth]{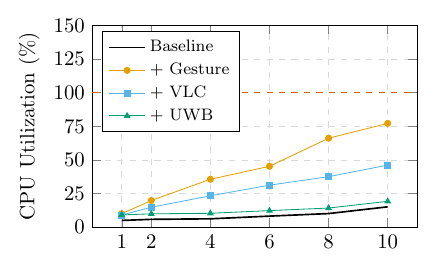}
    \captionsetup{font=footnotesize}
    \vspace{-5pt}
    \caption{CPU Utilization}
    \label{fig:cpu_overhead}
\end{subfigure}

\vspace{0.5em}

% Subfigure for GPU Utilization
\begin{subfigure}{\columnwidth}
    \centering
    \includegraphics[width=0.8\columnwidth]{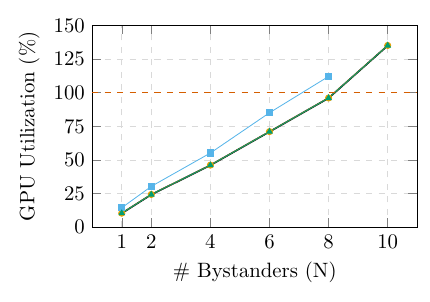}
    \captionsetup{font=footnotesize}
    \vspace{-10pt}
    \caption{GPU Utilization}
    \label{fig:gpu_overhead}
\end{subfigure}

\captionsetup{font=footnotesize}
\caption{CPU and GPU utilization as a function of the number of bystanders. Each data point was the average of 2 minutes with signaling every 15 seconds. In GPU utilization, the baseline, Gesture, and UWB are overlapping due to the lack of any additional GPU operations in these modalities. Any value above 100\% leads to frame dropping due to the fact that GPU usage is calculated based on inference time / 33.33ms (30 FPS).}
\label{fig:overhead}
\vspace{-1.5em}
\end{figure}

The results (Figure~\ref{fig:overhead}) show the system is fundamentally GPU-bound by the face detection pipeline. The baseline GPU load scales linearly with the number of faces, crossing the 100\% utilization threshold at N=8.
This establishes a hard architectural limit on the system's real-time scalability. The Hand Gesture and UWB modalities add negligible GPU load. The VLC modality's cost, in contrast, increases linearly with the number of on-screen search regions. The CPU utilization reflects the cost of the signaling logic. The baseline and UWB modality impose a near-constant, minimal CPU load. Conversely, the Hand Gesture modality's CPU cost scales steeply, a direct result of running a second vision model for hand tracking and per-user state management. This analysis confirms that while the GPU-bound face pipeline dictates system throughput, the CPU cost reveals a clear scalability trade-off: the UWB modality is more efficient at scale than its vision-based counterparts (gesture, LED).

\subsubsection{UWB Tag Power Consumption}
To validate the energy savings of our hybrid BLE+UWB protocol, we measured the tag's average power consumption using a Makerfire USB power meter. We considered  three separate 5-minute scenarios: (1) a baseline idle state, (2) a worst-case of continuous UWB ranging, and (3) typical signaling with our hybrid protocol. The results confirm our design's efficiency. Continuous UWB ranging averaged a power draw of 67.4\,mW. In contrast, our hybrid protocol averaged only 4.2\,mW, which is over \textbf{16$\times$ lower} and only marginally higher than the idle baseline of 0.5\,mW. This efficiency is a direct result of our design; the power-intensive UWB localization was required only twice during the test, with all subsequent signals handled by the energy-efficient `Fast Path'.

\subsection{Comparison to State-of-the-Art}
\label{sec:eval-comparison}

We position \tool{} against relevant prior work. A direct component-wise benchmark for the gesture and UWB modalities is not possible, as our work is the first to evaluate them for this task. We therefore structure our comparison in two parts. First, we benchmark our improved VLC modality against our prior work, BystandARIA~\cite{alaaraj2025bystandaria}. Second, we compare the bystander-centric control model of \tool{} against BystandAR~\cite{corbett2023bystandar}, a user-centric system.

\subsubsection{Comparison to Prior VLC Signaling}
The VLC modality in \tool{} is a significant improvement over our preliminary prior work, BystandARIA~\cite{alaaraj2025bystandaria}. BystandARIA achieved an average detection accuracy of 85\% across lighting conditions, dropping to 80\% at its maximum tested range of 4 meters. In addition, it processed the video stream at a rate of 20 frames per second (FPS). In contrast, \tool{}'s GPU-accelerated filter and Viterbi-based decoding achieve an accuracy of 85\% at up to 10 meters and 30 FPS. This demonstrates a substantial improvement in both robustness and real-time performance over the prior state-of-the-art.

\subsubsection{Comparison to Real-time Bystander Privacy Systems}
The fundamental difference between \tool{} and systems like BystandAR~\cite{corbett2023bystandar} lies in the control model. BystandAR is user-centric, inferring bystander status from the device wearer's gaze and speech. In contrast, \tool{} provides bystanders with direct, explicit control, closing this agency gap. This architectural divergence extends to hardware dependency and evaluation scope. BystandAR requires a HoloLens 2 with dedicated eye-tracking and depth sensors, while \tool{} operates on commodity smartphones. Consequently, BystandAR's evaluation reports a 98.1\% protection rate in a two-person, close-range (2m) scenario. Our evaluation characterizes performance up to 10m with up to six concurrent bystanders, demonstrating effectiveness at scale and range not addressed by prior user-centric systems.

Furthermore, \tool{} achieves this performance on more accessible hardware. BystandAR requires a HoloLens 2 and specialized sensors, including eye-tracking, SLAM, and depth sensors. \tool{} aoperates on a stock smartphone with only its standard camera and basic RF hardware. Despite this, its marginal CPU cost for the gesture (+29.4\% at 4 people) and UWB (+4.1\% at 4 people) modalities is comparable to or lower than BystandAR's (+27\%), proving the viability of a bystander-centric model on ubiquitous devices.

This shift in control model introduces a trade-off between latency and robustness. BystandAR's low latency stems from its reliance on instantaneous, implicit user gaze. \tool{}'s higher latencies are a consequence of decoding an explicit signal over time (gestures/VLC) or completing a ranging protocol (UWB), prioritizing signal integrity and security over instantaneous response. \tool{} thus establishes a new and complementary design point, proving that shifting control to the bystander is feasible without sacrificing real-time performance or requiring specialized hardware.

%% file: sections/discussion.tex
\section{Discussion}
\label{sec:discussion}

\subsection{Comparing Modalities}

Our evaluation demonstrates that no single signaling modality is optimal across all conditions. Instead, the choice of modality presents a trade-off between performance metrics like latency and range, and human-centric factors like comfort, convenience, cost, and availability.  The results in Table~\ref{tab:performance} and Figure~\ref{fig:overhead} highlight the strengths and weaknesses of the three modalities.

\subsubsection{Hand Gesture: Zero-Cost but High Overhead.}
The gesture modality's primary advantage is requiring no hardware. It achieves low latency ($\sim$150-200\,ms) and high accuracy (100\% at 1-2m) at close range. However, its operational range is limited to approximately 3 meters, beyond which hand landmarks cannot be reliably detected. It is also the only modality with a non-zero false positive rate (up to 9.17\% with multiple moving users), as interpreting human motion is inherently more ambiguous than decoding a structured digital signal. Critically, as shown in Figure~\ref{fig:cpu_overhead}, this modality incurs the highest marginal CPU cost, with utilization scaling steeply as more users are tracked. This is because the hand detection model runs on the CPU to avoid contention with the face detector on the GPU.

\subsubsection{VLC: Balanced but Environmentally Sensitive.}
The LED beacon strikes a balance between the other two modalities, offering a greater range than gestures (up to 10m) with a moderate latency of $\sim$650\,ms. Its use of a checksum-verified packet ensures a zero false positive rate. However, its performance is fundamentally limited by environmental factors. As shown in Table~\ref{tab:performance}, its accuracy degrades significantly in bright light, which is due to the bright surroundings being mistaken for LEDs. It also carries a computational cost, adding a moderate, linear overhead to both CPU and GPU as it processes more search regions for more users (Figure~\ref{fig:overhead}).

\subsubsection{UWB: Robust and Scalable at the Cost of Latency.}
The UWB modality prioritizes robustness and discretion. As an RF-based method, it is immune to all lighting conditions and line-of-sight occlusions. Its explicit GATT-based signal and physical-layer security properties result in a zero false positive rate and high accuracy. This comes at the cost of high latency ($\sim$2000\,ms), a deliberate design choice dominated by the burst-ranging process needed for a high-confidence spatial localization. A key advantage is its scalability. Its latency is nearly constant regardless of the number of users, and as Figure~\ref{fig:overhead} confirms, it has a negligible computational cost, making it the most efficient modality at scale.

\subsubsection{Recommendations}
Our evaluation suggests distinct use cases for each modality based on their performance trade-offs. The \textbf{Hand Gesture} modality is best suited for spontaneous, close-range interactions where bystanders lack hardware and computational overhead is permissible, such as in small office meetings. The \textbf{LED Beacon} is ideal for medium-range indoor environments with controlled lighting and a moderate number of users, like presentations or lectures. Finally, the \textbf{UWB Tag} is best for scenarios requiring discretion, robustness to visual conditions, or scalability to crowds, particularly when reliability is prioritized over latency.

\subsection{Limitations and Future Work}
\label{sec:limitations}

\noindent\textbf{Hardware and Deployment.}
Our implementation relies on custom-programmed tags for the VLC and UWB modalities. A path to deployment involves transitioning to commodity hardware. For UWB, this means leveraging the native support now integrated into modern smartphones, which would avoid the need for external tags. For VLC, signaling could be integrated into future wearables like smart glasses or use existing hardware, such as a smartphone's flashlight LED. Beyond hardware, a significant hurdle for wide-scale adoption is the lack of a standardized discovery and communication protocol. Future work should focus on creating such a standard to allow any user's device to seamlessly interact with any \tool{}-enabled camera system.

\noindent\textbf{Adversarial Scenarios.}
Our evaluation focused on system performance and robustness against adversarial or incidental interference, such as accidental motion or occlusions in a crowded scene. Our signal-source validation provides a foundational defense against impersonation by requiring geometric consistency. For instance, it ensures the distance measured by UWB matches the distance estimated from the face bounding box to within a 10\% tolerance, rejecting signals that fail this check. While this provides a strong baseline against impersonation, future work could further investigate the system's resilience against sophisticated, active adversaries. This includes evaluating the UWB and VLC channels against radio-level spoofing and replay attacks.

\noindent\textbf{Face Pipeline Constraints.}
The underlying face pipeline imposes two fundamental constraints. First, GPU-based face detection limits real-time throughput to approximately eight concurrent bystanders, beyond which processing time exceeds the frame budget. Future work could mitigate this by offloading inference to dedicated NPUs or using intermittent detection with lightweight trackers. Second, the pipeline dictates the system's 10-meter effective range, a direct consequence of the minimum pixel resolution required by the face detector. Improvements in long-range, low-resolution object detection would directly extend the operational range of all vision-based modalities.

\noindent\textbf{User Study.}
We conducted a preliminary user study to assess usability and preference; the methodology, results, and participant responses are available in Appendix~\ref{sec:appendix-user-study}. The study revealed a strong user preference for the subtle, app-based UWB modality over the more conspicuous gesture and LED methods. Future work should validate these preliminary findings through a large-scale study to better understand broader user opinions on such on-device privacy systems.

\noindent\textbf{Possible Extensions.}
The modular design of \tool{} allows for several extensions. First, while this paper focused on faces, the pipeline can be adapted to detect and anonymize other sensitive information within a video stream, such as vehicle license plates. Second, the design of each signaling modality can be improved. The architecture supports incorporating alternative gestures or designing different communication protocols on top of the VLC or UWB physical layers to meet different application requirements.
Third, \tool{} can  incorporate new and improved ``face pipelines'' for face detection and blurring, including  emerging ones based on {\bf visual language models (VLMs)}. 
 VLMs may infer identity from contextual cues (e.g., clothing, body shape, environment) that remain in the anonymized frame. Future work must evaluate the robustness of redaction methods against VLM-based re-identification attacks. Conversely, VLMs could also enable more advanced, context-aware privacy controls, moving beyond simple redaction to semantic modification of the scene. \tool{}  could leverage a lightweight VLM to interpret more nuanced privacy intentions directly from the visual context, moving beyond explicit, predefined signals.

\subsection{Ethical Considerations} \label{sec:ethics}

The design and evaluation of \tool{} were conducted in accordance with our university's IRB guidelines. Following consultation with an IRB specialist, this project was determined to be eligible for exempt self-determination. This is because our system does not collect or store any personally identifiable information (PII). All data processing is performed on-device for the sole purpose of evaluating the system's real-time signaling and anonymization capabilities.

%% file: sections/conclusion.tex
\section{Conclusion}
\label{sec:conclusion}
This paper addressed the critical gap in bystander privacy control for mobile cameras. We presented \tool{}, a complete, on-device system that enables individuals to signal their privacy preferences in real-time. We designed, implemented, and evaluated three novel signaling modalities: a device-free hand gesture, a hardware-based VLC beacon, and a secure UWB tag. Our empirical evaluation quantified the distinct performance trade-offs of each method. The gesture offers low latency at short range, the VLC beacon provides a balanced solution for indoor environments, and the UWB tag delivers robust, discreet signaling at the cost of higher latency. By demonstrating the viability of these complementary channels, unified by a common signal-source validation framework, this work provides the first comparative analysis of practical signaling techniques for bystander privacy and offers the  architectural foundation for bystander-controlled privacy in camera-enabled systems.

%% file: sections/acks.tex
\section*{Acknowledgment}
This work was supported in part by the National Science Foundation under award number 1956393. We would like to thank Salma Elmalaki and Aditi Majumder for their input.

%% file: sections/appendix.tex
\appendix

\section{System Design Details}
This section provides supplementary implementation details for the system architecture presented in Section~\ref{sec:design}. We elaborate on the pipeline stages for each of the three signaling modalities.

\subsection{Selection of Processing Resolution}\label{app:proc_resolution}
We select a processing resolution of 1280x960 (1.2MP). This resolution represents a deliberate trade-off between maximizing face detection range and meeting the 30 FPS processing target on a mobile System-on-Chip (SoC). Higher resolutions introduce prohibitive computational and memory bandwidth costs for real-time operation. Conversely, lower resolutions (e.g., 640x480) unacceptably reduce the system's effective range. The 4:3 aspect ratio also matches the native format of many mobile camera sensors, avoiding performance penalties from on-the-fly cropping or distortion. This principle of selecting a task-specific resolution over the maximum available sensor resolution is standard in resource-constrained vision systems where balancing latency and effective range is critical, such as in automotive platforms (e.g. self driving cars such as Tesla's HW3 Cameras also use the same ratio and resolution).

\subsection{Hand Model Selection}\label{app:hand-model}
The gesture modality requires a robust, real-time hand keypoint detector. We evaluated two candidate algorithms: Google's MediaPipe HandLandMarker and a YOLOv11n-pose model. The latter was trained on the public `hand-keypoints' dataset. Our evaluation focused on the maximum operational distance at which each model could reliably perform keypoint detection. MediaPipe consistently provided robust detection up to a distance of 3 meters. In contrast, the YOLOv11n-pose model's effective range was limited to 2.5 meters. Given its superior range and comparable inference latency, we selected MediaPipe Hands for the gesture modality in \tool{}.

\subsection{Camera Configuration for Signal Acquisition.}\label{app:camera-config}
Reliable VLC decoding requires a stable and predictable camera stream. The automatic exposure and gain control algorithms common in mobile cameras are detrimental to this task, as they can artificially brighten the LED's `off' state or dim its `on' state, destroying the signal's dynamic range. To counteract this, our system leverages the Camera2 API to disable all automatic exposure controls (\texttt{\seqsplit{CONTROL\_AE\_MODE\_OFF}}). We enforce a fixed, short exposure time (\texttt{\seqsplit{SENSOR\_EXPOSURE\_TIME}}) and a high sensor sensitivity (\texttt{\seqsplit{SENSOR\_SENSITIVITY}}). This configuration ensures that the LED's blinking pattern is captured with high fidelity, producing sharp temporal transitions essential for the decoding logic.

To ensure a high-fidelity optical signal, the camera's automatic exposure and gain controls must be disabled. These features, designed for photography, corrupt the signal by altering the perceived brightness of the LED's `on' and `off' states. To counteract this, our system leverages the Camera2 API to disable all automatic exposure controls (\texttt{\seqsplit{CONTROL\_AE\_MODE\_OFF}}). We enforce a fixed exposure time and sensor sensitivity based on ambient light conditions. This configuration ensures the LED's blinking pattern is captured with sharp temporal transitions essential for decoding.

\subsection{UWB Face Expansion Factor}\label{app:uwb-face-factor}
This factor is derived from anthropometric data showing the average shoulder breadth is approximately 2.5 times the head breadth (using the female biacromial-to-head-breadth ratio of $\frac{biacromialBreadth}{headBreadth} = \frac{14.28}{5.82} = 2.47$~\cite{gordon2014anthropometric}). This creates a robust binding area that accounts for natural variations in tag placement relative to the user's face.

\section{Evaluation Details}
This section provides supplementary details for the evaluation presented in Section~\ref{sec:evaluation}. We describe the characterization of the distance estimation models used for the geometric consistency checks of the hand gesture and VLC modalities.

\subsection{Modality 1: Hand Distance Estimation}\label{app:hand-size}
\begin{figure}[t]
\centering
\includegraphics[width=0.9\columnwidth]{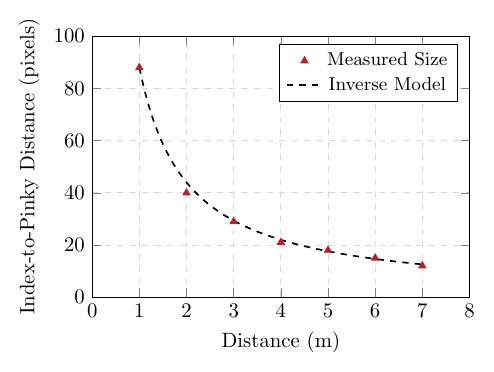}
\captionsetup{font={footnotesize}}
\caption{On-screen distance between index and pinky keypoints as a function of subject distance. Measurements beyond 3\,m were enabled by MediaPipe's tracking mode after an initial close-range detection.}
\label{fig:hand_distance}
\vspace{-1.5em}
\end{figure}

To defend against impersonation attacks where a malicious actor's hand is paired with the face of a distant victim bystander, \tool{} performs a geometric consistency check. The system independently estimates the distance to a face and a gesturing hand using their respective on-screen sizes. For the hand, we first characterized the relationship between a user's distance and the pixel separation of their index and pinky fingertips, a stable proxy for hand size. As shown in Figure~\ref{fig:hand_distance}, the measured keypoint distance follows a predictable inverse model, decreasing from 88\,pixels at 1\,m to 12\,pixels at 7\,m. This data was collected by leveraging MediaPipe's tracking mode beyond its 3\,m initial detection limit. \tool{} accepts a gesture signal only if the distance inferred from the hand's size is within a tolerance threshold of the distance inferred from the associated face's bounding box (Figure~\ref{fig:distance_characterization}), ensuring both are geometrically consistent.

\subsection{Modality 2: LED Distance Estimation}\label{app:led-size}
\begin{figure}[t]
\centering
\includegraphics[width=0.9\columnwidth]{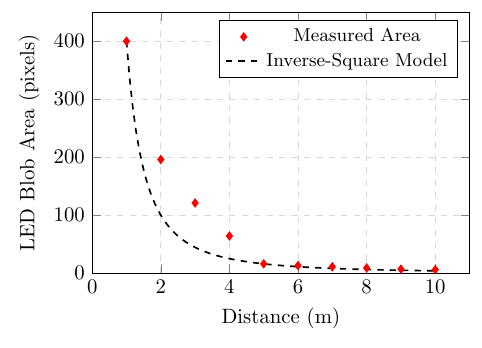}
\captionsetup{font={footnotesize}}
\caption{On-screen pixel area of the VLC beacon's LED as a function of distance. The data follows an inverse-square model, enabling geometric validation.}
\label{fig:vlc_blob_distance}
\vspace{-1.5em}
\end{figure}

To defend against impersonation attacks, \tool{} validates that the VLC signal originates from the same distance as the user's face (Figure~\ref{fig:distance_characterization}). This is achieved by comparing the distance inferred from the face's bounding box with the distance inferred from the on-screen pixel area (blob size) of the beacon's LED. We characterized the relationship between blob area and distance, as shown in Figure~\ref{fig:vlc_blob_distance}. The measured area follows the expected inverse-square model, decreasing from \texttt{400}\,pixels at 1\,m to an estimated \texttt{4}\,pixels at 10\,m. A decoded VLC message is accepted only if the distance inferred from its blob size is geometrically consistent with the distance inferred from the associated face.

\section{User Study Details}
\label{sec:appendix-user-study}

To evaluate usability, social acceptability, and user preference, we conducted a preliminary user study with 5 participants. After successfully using each of the three modalities, participants took part in a semi-structured interview. The following subsections detail the questionnaire used, a summary of the findings, and the detailed participant responses. This extends Section~\ref{sec:discussion}.

\subsection{Questionnaire}
\label{sec:appendix-questions}
The following questions were used to assess usability, social acceptability, trust, and overall user preference. Where noted, questions were accompanied by a 5-point Likert scale.

\subsection{Part 1: Per-Modality Feedback}
These questions were asked immediately after the user successfully tested a specific modality.

\subsubsection{Modality 1: Hand Gesture}
\begin{enumerate}
    \item \textbf{Learnability \& Usability:} On a scale of 1 (Very Difficult) to 5 (Very Easy), how easy or difficult was it to perform the correct gesture to activate the system? What made it easy or difficult?
    \item \textbf{Social Acceptability:} How comfortable would you feel making this gesture in a real public setting (e.g., a cafe)? Would you feel self-conscious? Why or why not?
    \item \textbf{Robustness \& Trust:} Did you ever feel the system might activate by accident while you were just talking or moving your hands normally? How much do you trust it to only work when you intend it to?
\end{enumerate}

\subsubsection{Modality 2: VLC Beacon}
\begin{enumerate}
    \item \textbf{Effort \& Convenience:} On a scale of 1 (Very Inconvenient) to 5 (Very Convenient), how convenient do you find the idea of carrying and using this small device?
    \item \textbf{Discreteness \& Subtlety:} How did using the LED feel compared to the hand gesture? Did it feel more private and subtle, or more obvious?
    \item \textbf{Mental Model:} Was it clear what you needed to do to make the beacon work?
\end{enumerate}

\subsubsection{Modality 3: UWB Tag}
\begin{enumerate}
    \item \textbf{Convenience \& Effort:} On a scale of 1 (Very Inconvenient) to 5 (Very Convenient), how do you rate using a simple app to control your privacy? How does the effort of using an app compare to performing a physical gesture or pointing a beacon?
    \item \textbf{Discreteness \& Social Comfort:} This method lets you signal without a visible action towards the camera. How important is this subtlety to you? Would you feel more comfortable using this app in public compared to the other methods?
    \item \textbf{Confidence \& Feedback:} Since the action happens on your own device, how confident were you that the camera system received your request? Did you feel a need to look at the recorder's screen to confirm your face was blurred?
\end{enumerate}

\subsection{Part 2: Comparative and Overall Feedback}
These questions were asked after the user had experienced all three modalities.

\begin{enumerate}
    \item \textbf{Overall Preference:} Please rank the three methods from your most preferred to your least preferred. Could you explain the reasons for your ranking?
    \item \textbf{Situational Use:} Can you imagine different situations where you would prefer one method over the others? (e.g., using the gesture when your hands are free, the beacon when sitting at a table, or the tag when walking through a crowd).
    \item \textbf{Problem \& Solution Value:} Overall, how valuable do you find the ability to control your privacy in this way? Is this a problem you have personally felt?
\end{enumerate}

\subsection{Summary of Results}
\label{sec:appendix-summary}

\begin{table*}[t]
\centering
\small
\begin{tabular}{|p{0.5cm}|p{2.5cm}|p{6cm}|p{6cm}|}
\hline
\textbf{P\#} & \textbf{Ranking (Gesture, LED, UWB)} & \textbf{Justification for Ranking} & \textbf{Situational Preference} \\
\hline
1 & 3, 2, 1 & ``I'd just much rather prefer the tag in every situation... I also just like a gesture free approach which is why I put the LED second.'' & ``If the UWB on the phone could act as a beacon I'd probably just use an app... Especially if my hands aren't free the quick LED signal is a bit more obvious, but way more convenient.'' \\
\hline
2 & 1, 3, 2 & (No direct justification given for ranking) & ``When hand not free, e.g. gym for badminton, prefer UWB. 2) walk in a crowd, UWB'' \\
\hline
3 & 3, 1, 2 & ``I liked the gesture in terms of usability as I don't have to carry anything... But in terms of privacy and data control, other two methods are better.'' & ``All the situation I would prefer gesture but if I am in a space where I want more privacy and control, I would prefer other two.'' \\
\hline
4 & 3, 2, 1 & ``I personally like less hand gesture. Compared with those devices, hand gesture feel not subtle.'' & ``In general, in a crowd, I feel the tag is more effective.'' \\
\hline
5 & 3, 1, 2 & ``I would prefer UWB. But I do not know whether it can handle several UWB at the same time.'' & ``If it can [handle multiple users], I would prefer UWB in every scenario.'' \\
\hline
\end{tabular}
\caption{Selected qualitative responses from the comparative user study questions.}
\end{table*}

\subsubsection{Quantitative Results}
When asked to rank the three methods, a strong preference emerged for the UWB modality, which was selected as the most preferred method by 4 out of 5 participants. The gesture modality was ranked first by one participant but last by two others, indicating a polarized reception. The LED modality was consistently ranked as the second or third choice and was never the most preferred. On a 5-point Likert scale, the gesture was rated highest for ease of use (avg. 4.6/5), while the LED (avg. 3.8/5) and UWB (avg. 3.8/5) modalities were rated highly for convenience.

\subsubsection{Qualitative Results}
The qualitative feedback reveals the trade-offs that informed user preferences.

\noindent\textbf{UWB Preference: Subtlety and Convenience.}
Participants who preferred the UWB modality cited its \textit{subtlety} as the primary benefit. The ability to signal without a visible action was highly valued, with one user noting it would be their choice in ``every situation'' if integrated into their phone. This aligns with the sentiment that ``Everyone's on their phones nowadays,'' making an app-based interaction feel natural and discreet, especially in crowded public spaces. The main drawback cited was the need to trust the ``hidden'' signal, with some users initially feeling a need to visually confirm that the system had received their request.

\noindent\textbf{Gesture Preference: Device-Free Interaction.}
The single participant who ranked the gesture first valued its ``device-free'' nature, requiring no extra hardware. Other participants agreed it was easy to learn and perform. However, the primary concerns for the gesture were the potential for social awkwardness and the risk of false positives from normal hand movements, with one user noting a ``moderate trust'' in its reliability.

\noindent\textbf{LED as a Middle Ground.}
The LED beacon was generally seen as a functional compromise. It was perceived as more private and subtle than the gesture but more obvious than the UWB tag. Its main appeal was as a simple alternative for situations where hands are not free and a gesture is inconvenient.

\noindent\textbf{Situational Use and Overall Value.}
A key finding was that participants could imagine situational uses for different modalities, validating the multi-modal design. For example, UWB was preferred for walking through a crowd or when hands are occupied (e.g., at the gym), while the gesture was acceptable when hands are free. All participants reported that the ability to control their privacy in this manner was ``super valuable'' or ``very valuable,'' confirming the importance of the core problem.

\subsection{Detailed Participant Responses}
\label{sec:appendix-responses}
The table below provides a selection of direct qualitative feedback from the comparative portion of the interview. The ranking is ordered (Most Preferred, Second, Least Preferred).